%% file: 60fe_thermalActivation.tex
\documentclass[twocolumn,showpacs,preprintnumbers,amsmath,amssymb,floatfix,superscriptaddress,PRC]{revtex4}
\usepackage{graphicx}
\usepackage{dcolumn}
\usepackage{bm}

\begin{document}
\title{The thermal neutron capture cross section of the radioactive isotope $^{60}$Fe}

\author{T.~Heftrich}
\email{heftrich@iap.uni-frankfurt.de}
\affiliation{Goethe-Universit\"{a}t Frankfurt, Frankfurt, Germany}
\author{M.~Bichler}
\affiliation{Technische Universit\"{a}t Wien, Vienna, Austria}
\author{R.~Dressler}
\affiliation{Paul-Scherrer-Institut, Villigen, Switzerland}
\author{K.~Eberhardt}
\affiliation{Johannes Gutenberg-Universit\"{a}t Mainz, Mainz, Germany}
\author{A.~Endres}
\affiliation{Goethe Universit\"{a}t Frankfurt, Frankfurt, Germany}
\author{J.~Glorius}
\affiliation{Goethe Universit\"{a}t Frankfurt, Frankfurt, Germany}
\affiliation{GSI Helmholtzzentrum f\"ur Schwerionenforschung, Darmstadt, Germany}
\author{K.~G\"{o}bel}
\affiliation{Goethe Universit\"{a}t Frankfurt, Frankfurt, Germany}
\author{G.~Hampel}
\affiliation{Johannes Gutenberg-Universit\"{a}t Mainz, Mainz, Germany}
\author{M.~Heftrich} 
\affiliation{Goethe Universit\"{a}t Frankfurt, Frankfurt, Germany}
\author{F.~K\"{a}ppeler}
\affiliation{Karlsruhe Institute of Technology, Karlsruhe, Germany}
\author{C.~Lederer}
\affiliation{University of Edinburgh, Edinburgh, UK}
\author{M.~Mikorski} 
\affiliation{Goethe Universit\"{a}t Frankfurt, Frankfurt, Germany}
\author{R.~Plag}
\affiliation{Goethe Universit\"{a}t Frankfurt, Frankfurt, Germany}
\author{R.~Reifarth} 
\affiliation{Goethe Universit\"{a}t Frankfurt, Frankfurt, Germany}
\author{C.~Stieghorst}
\affiliation{Johannes Gutenberg-Universit\"{a}t Mainz, Mainz, Germany}
\author{S.~Schmidt}
\affiliation{Goethe Universit\"{a}t Frankfurt, Frankfurt, Germany}
\author{D.~Schumann}
\affiliation{Paul-Scherrer-Institut, Villigen, Switzerland}
\author{Z.~Slavkovsk\'a}
\affiliation{Goethe Universit\"{a}t Frankfurt, Frankfurt, Germany}
\author{K.~Sonnabend} 
\affiliation{Goethe Universit\"{a}t Frankfurt, Frankfurt, Germany}
\author{A.~Wallner} 
\affiliation{Australian National University, Canberra, Australia}
\author{M.~Weigand}
\affiliation{Goethe Universit\"{a}t Frankfurt, Frankfurt, Germany}
\author{N.~Wiehl}
\affiliation{Johannes Gutenberg-Universit\"{a}t Mainz, Mainz, Germany}
\author{S.~Zauner}
\affiliation{Johannes Gutenberg-Universit\"{a}t Mainz, Mainz, Germany}

\date{\today}
\begin{abstract}
\textbf{Background:} 50\,\% of the heavy element abundances are produced via slow neutron capture reactions in different stellar scenarios. The underlying nucleosynthesis models need the input of neutron capture cross sections. \\
\textbf{Purpose:} One of the fundamental signatures for active nucleosynthesis in our galaxy is the observation of long-lived radioactive isotopes, such as $^{60}$Fe with a half-life of $2.60\times10^6$\,yr. To reproduce this $\gamma$-activity in the universe, the nucleosynthesis of $^{60}$Fe has to be understood reliably. \\
\textbf{Methods:} A $^{60}$Fe sample produced at the Paul-Scherrer-Institut was activated with thermal and epithermal neutrons at the research reactor at the Johannes Gutenberg-Universit\"{a}t Mainz. \\
\textbf{Results:} The thermal neutron capture cross section has been measured for the first time to $\sigma_{\text{th}}\,=\,0.226 \ (^{+0.044}_{-0.049}) \ \text{b} $. An upper limit of $\sigma_{\text{RI}} < 0.50\ \text{b}$ could be determined for the resonance
integral.\\
\textbf{Conclusions:} An extrapolation towards the astrophysicaly interesting energy regime between \mbox{$kT$\,=\,10 keV} and 100\,keV illustrates that the s-wave part of the direct capture component can be neglected.
\end{abstract}

\maketitle

\section{Introduction}
The decays of the unstable isotopes $^{60}$Fe \mbox{($t_{1/2} = 2.60$\,Myr \cite{WBB15})} and 
$^{26}$Al \mbox{($t_{1/2} = 0.717$\,Myr \cite{SWE72})} in the Milky Way, which have been 
observed with satellite-based $\gamma$-ray telescopes \cite{Smi03,HKJ05}, are considered
as a clear signature of ongoing stellar nucleosynthesis \cite{TWH95}.   

\begin{figure}[t!]
\includegraphics[width=0.49\textwidth]{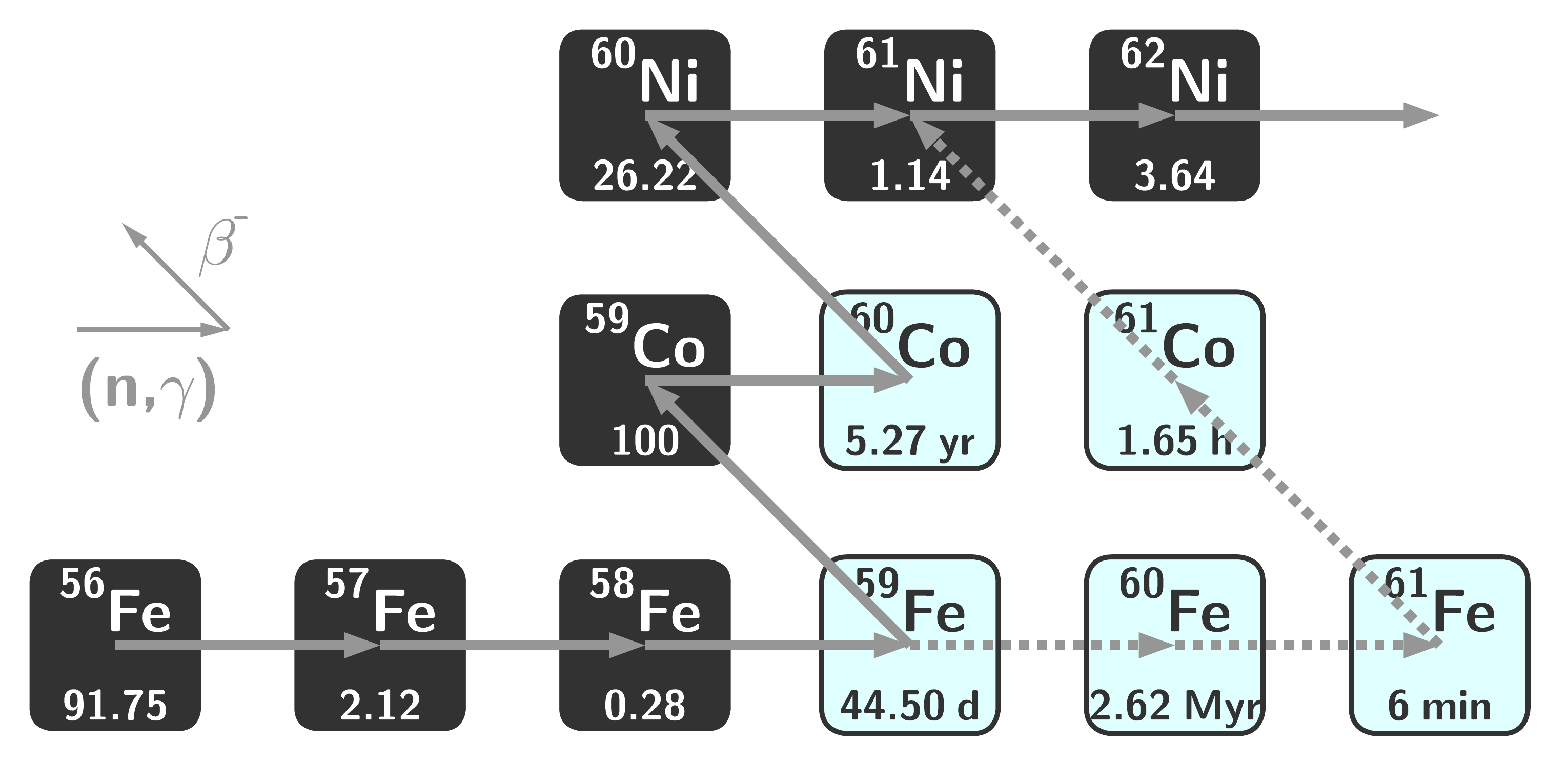}
\caption{(Color online) The $s$-process reaction path between Fe and Ni. The isotope $^{60}$Fe is produced 
via a sequence of (n,$\gamma$) reactions starting at the stable iron isotopes. Because of the 
short half-life of $^{59}$Fe ($t_{1/2}~\approx~45$ d), the production of $^{60}$Fe depends 
critically on the stellar neutron density.}  
\label{path}
\end{figure}

The production of $^{60}$Fe in the slow neutron capture process ($s$-process)  \cite{TWH95}
is hampered by the rather short-lived precursor $^{59}$Fe ($t_{1/2}\,=\,44.495$\,d \cite{Bag02}), which 
acts as a branch point of the $s$-process path as illustrated in Fig.~\ref{path}. Accordingly,
high neutron densities are required to avoid that the reaction flow is bypasses $^{60}$Fe
via the decay of $^{59}$Fe. Once $^{60}$Fe is reached, it can also be destroyed 
by neutron capture or - on longer time scales - by $\beta^{-}$-decay. High neutron densities 
are generally accompanied by very high temperatures, but the synthesis of $^{60}$Fe requires 
an upper limit of about \mbox{$2 \times 10^9$ K} ($T_{9}$ = 2), because photodisintegration 
reactions such as \mbox{$^{60}$Fe($\gamma, \text{n}$)} and $^{59}$Fe($\gamma, \text{n}$) start to dominate
otherwise. 

\begin{figure*}[t] 
\includegraphics[width=0.99\textwidth]{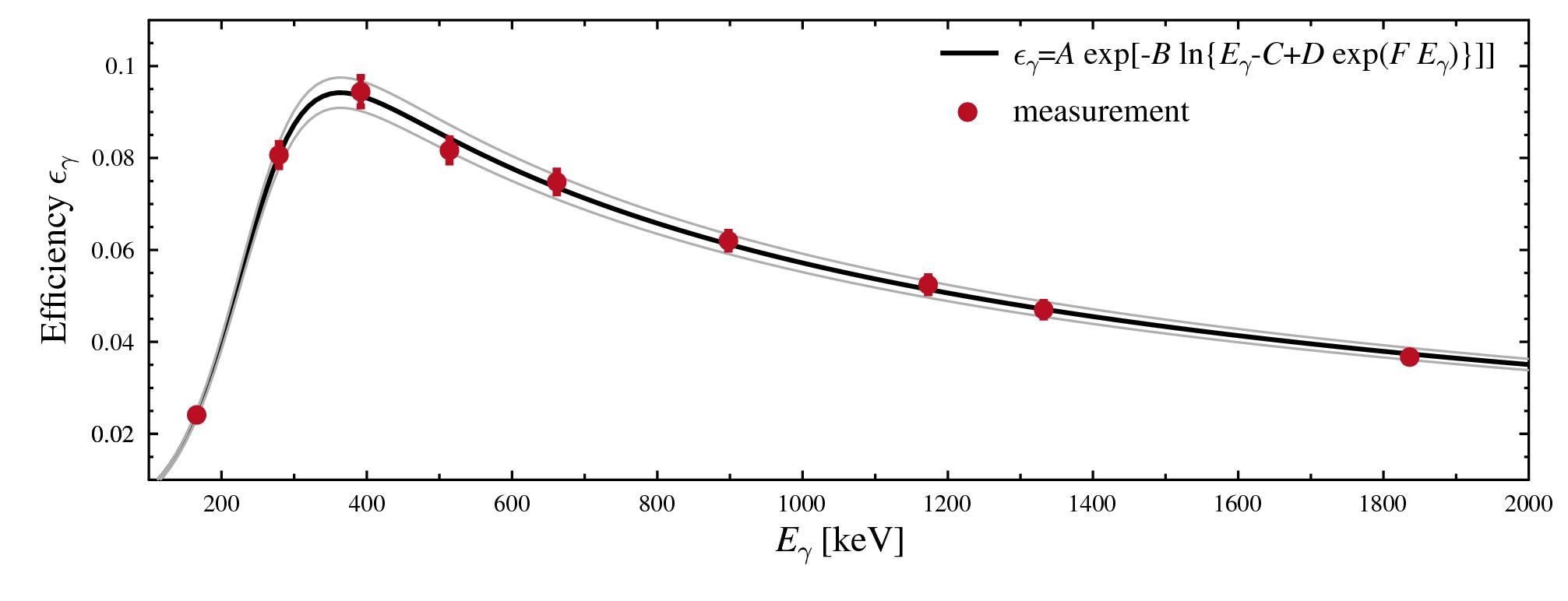}
\caption{(Color online) The detection efficiency of the HPGe detector used for the activity measurements in Mainz. The efficiency used for determination the number of $^{60}$Fe particles is given in \mbox{Table \ref{CalStandard}}. 
The solid line shows the least squares fit to interpolate between the data points of the calibrated solution (red). For $^{88}$Y and $^{60}$Co the data points were corrected for cascade effects using the GEANT-3.21 package \cite{GEA93,DPA04}. The grey band represents the uncertainty of the fit.}
\label{xleff}
\end{figure*}

There are two different astrophysical scenarios where $^{60}$Fe can be produced \cite{LCh06}: 
during the He-shell burning phase in low-mass thermally pulsing asymptotic giant branch (AGB) 
stars and during the convective \mbox{C-shell} burning in massive pre-supernova stars. In AGB stars, neutron densities 
of \mbox{10$^{10}$ cm$^{-3}$} and temperatures around \mbox{$2.5 \times 10^8$ K} ($T_8=~2.5$) are reached, 
whereas in massive stars neutron densities of \mbox{10$^{12}$ cm$^{-3}$} at temperatures 
of up to $T_8 = 10$ during C-shell burning are reached \cite{RLK14}. According to detailed 
stellar model calculations by Limongi and Chieffi \cite{LCh06}, about 65\,\% of the total yield of 
$^{60}$Fe are in fact synthesized in the pre-supernova stage of massive stars and 18\,\% are 
contributed by the He burning shell of less massive stars. A third major component is eventually 
produced by explosive shell burning during the supernova itself. These contributions to the total 
$^{60}$Fe yield are strongly affected by the respective masses and metallicities of the stars 
involved and may vary correspondingly.

A crucial input for the production of $^{60}$Fe in AGB stars and massive pre-supernova stars are 
the neutron capture cross sections at the respective stellar temperatures. So far, an activation 
measurement of the $^{60}$Fe($\text{n}, \gamma$)$^{61}$Fe cross section at neutron energies 
corresponding to a thermal energy of $kT$~=~25\,keV (typical for AGB stars) was performed 
at Forschungszentrum Karlsruhe, Germany. The Maxwellian averaged cross section (MACS) at 
$kT$~=~30\,keV was determined to ($5.15\pm1.4$)~mb \cite{URS09}. The direct 
capture (DC) component of the cross section at this temperature constitutes an important 
information for the extrapolation towards the astrophysically interesting temperatures in 
massive stars around $kT$~=~90\,keV. In this respect, the thermal cross section provides a constraint for the s-wave component of the DC cross section. Therefore, the previously unknown thermal cross section of $^{60}$Fe was measured using the irradiation facility at the TRIGA (Training, Research, Isotopes, General Atomic) type research reactor at Johannes Gutenberg-Universit\"{a}t Mainz, Germany \cite{EbK00,HET06}.

 \section{Experiment}
The $^{60}$Fe sample was produced at the Paul-Scherrer-Insitut (PSI) in Villigen, 
Switzerland \cite{SND10}. In order to compensate for the limited amount of $^{60}$Fe the only possible method 
for the determination of the neutron capture cross section was an integral activation measurement at high neutron fluxes. Compared to the more generally applicable time-of-flight technique, the activation method has the 
advantage of excellent sensitivity \cite{RLK14}, which allows neutron capture measurements 
even on very small samples \cite{RAH03,ReK02}. This technique has the additional advantage 
that it does not require isotopically enriched samples, because the capture reactions can be 
identified via the $\gamma$-decay characteristics \mbox{of the product nucleus $^{61}$Fe.}
\begin{table*}[t]
\centering  
   \caption{Decay characteristics and efficiency of $\gamma$-emission of the investigated nuclei.}
   \begin{ruledtabular}
   \begin{tabular}{l l l l l l}
 Isotope 		& $t_{1/2}$  			& E$_\gamma$	/keV                 &  I$_\gamma$ /\,\% 	& Reference		&$\epsilon_{\gamma}$	\\
            \hline 
$^{60}$Co 	& ($1925.28\pm0.14$) d    & $1173.228\pm 0.003$ 		& $99.85\pm0.03$ 	& \cite{Bro13}   & $0.024\pm 0.0002^{a}$\\
 			& 			         	& $1332.492\pm 0.004$	        & $99.9826\pm0.0006$ 	& 			&  $0.022\pm 0.0002^{a}$	\\
$^{97}$Zr 	& ($16.749\pm0.008$) h 	& $743.36\pm 0.03$ 	        & $93.09\pm 0.16$ 	        & \cite{Nic10} 	&  $0.069\pm 0.002^{b}$ \\
$^{95}$Zr 	& ($64.032\pm0.006$) d 	& $724.195\pm 0.004$ 		& $44.27\pm 0.22$ 	        & \cite{BMS10}  & $0.070\pm 0.001^{b}$	\\
 			& 					& $756.728\pm 0.012$ 		& $54.38\pm 0.22$ 	        & 			& $0.068\pm 0.002^{b}$	\\
$^{61}$Fe 	& ($5.98\pm0.06$) min 	& $297.90\pm 0.07$ 	        	& $22.24\pm2.88$         	& \cite{Bha99} 	&  $0.087\pm 0.002^{b}$\\
 			& 					& $1027.42\pm 0.11$ 	        	& $42.73\pm4.92$ 	        & 			& $0.056\pm 0.002^{b}$	\\
			& 					& $1205.07\pm 0.12$    		& $43.60\pm4.50$ 	        & 			& $0.051\pm 0.001^{b}$	\\
\end{tabular}
\end{ruledtabular}
 \label{CalStandard}
$ ^{a}$ measured with a HPGe detector at Goethe-Universit\"{a}t Frankfurt (used for 60Fe determination).\\
 $^{b}$ measured with a different HPGe detector at the research reactor at Johannes Gutenberg-Universit\"{a}t Mainz  (used for $^{61}$Fe and Zirconium determination)
\end{table*}

\subsection{Measurements and calibration}
The induced activities were measured using a HPGe detector (CANBERRA-GX7020) with a relative efficiency of 72.3\,\%. The output signals from the preamplifier were converted with a flash-ADC (CAEN module V1724). The dead time corrections were determined using a $^{137}$Cs sample, which was placed at a fixed distant position during all activity measurements. The corresponding corrections were negligibly small. Because of the contamination of $^{55}$Fe in the $^{60}$Fe sample, the activity of $^{55}$Fe was suppressed by a lead foil 1\,mm in thickness. 

The efficiency was determined by a calibrated solution containing the standard single- or double-line $\gamma$-ray emitters $^{60}$Co, $^{85}$Sr, $^{88}$Y, $^{113}$Sn, $^{137}$Cs, $^{139}$Ce, and $^{203}$Hg. The uncertainty of the $\gamma$-emission rate was given with 3\,\% (2$\sigma$). This multi-nuclide solution was absorbed in a pure graphite disc 6\,mm in diameter and 1\,mm in thickness to match the properties of the $^{60}$Fe sample used in the measurement (see below). For all $\gamma$-activity measurements, the samples were placed 7.4\,mm in front of the Ge crystal. Because of the small distance between sample and detector, cascade corrections were necessary for the decays of $^{60}$Co and $^{89}$Y. Those corrections were based on the simulations performed using the \mbox{GEANT-3.21 package} \cite{GEA93,DPA04}. The corresponding correction for the emission line of $^{60}$Co at the energy of \mbox{1173\,keV }was 30\,\%, at \mbox{1332\,keV} 31\,\% and for $^{88}$Y at the energy of \mbox{898\,keV} and \mbox{1836\,keV} 27\,\% and 29\,\%, respectively.  As shown in Fig.~\ref{xleff} the measured efficiencies could be reproduced within the experimental uncertainties of $\pm3.5$\,\% over the energy range from 150\,keV to 1900\,keV by the expression
\begin{equation}
\epsilon_{\gamma}=A\ {\rm exp}{\Large [}-B\ \ln \{ E_{\gamma}-C+D\ {\rm exp}({F\times E_{\gamma}) \} {\Large ]}}.
\end{equation}

\begin{figure}[b]
\includegraphics[width=0.45\textwidth]{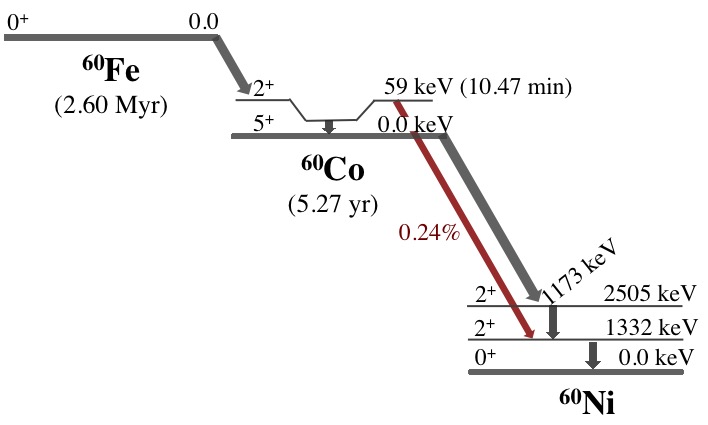}
\caption{(Color online) The activity of the $^{60}$Fe sample was determined by the $\gamma$-ray cascade with energies of 1173\,keV and 1332\,keV emitted after the $\beta$-decay of the daughter nucleus $^{60}$Co. Data from \cite{RFK09,Bro13}.}  
\label{decay60fe}
\end{figure}

\subsection{Sample preparation}\label{SamplePreparation}
The $^{60}$Fe was extracted from slices of a cylindrical copper beam dump, which was previously 
irradiated with 590\,MeV protons at PSI \cite{SND10}. In addition to $^{60}$Fe activity, the initial 
copper sample of 3\,g also contained 150\,MBq of $^{60}$Co, 100\,MBq of $^{55}$Fe, and 2\,MBq 
of $^{44}$Ti. Details of the chemical separation of the $^{60}$Fe fraction are described in \cite{SND10}. The final purification was performed shortly before the experiment using liquid-liquid extraction into methyl-isobutyl ketone from 7 M HCl solution and following back-extraction with diluted HCl. This solution was evaporated on a graphite disc with 6 mm diameter and 1 mm thickness. 
\begin{figure}[b]
\includegraphics[width=0.5\textwidth]{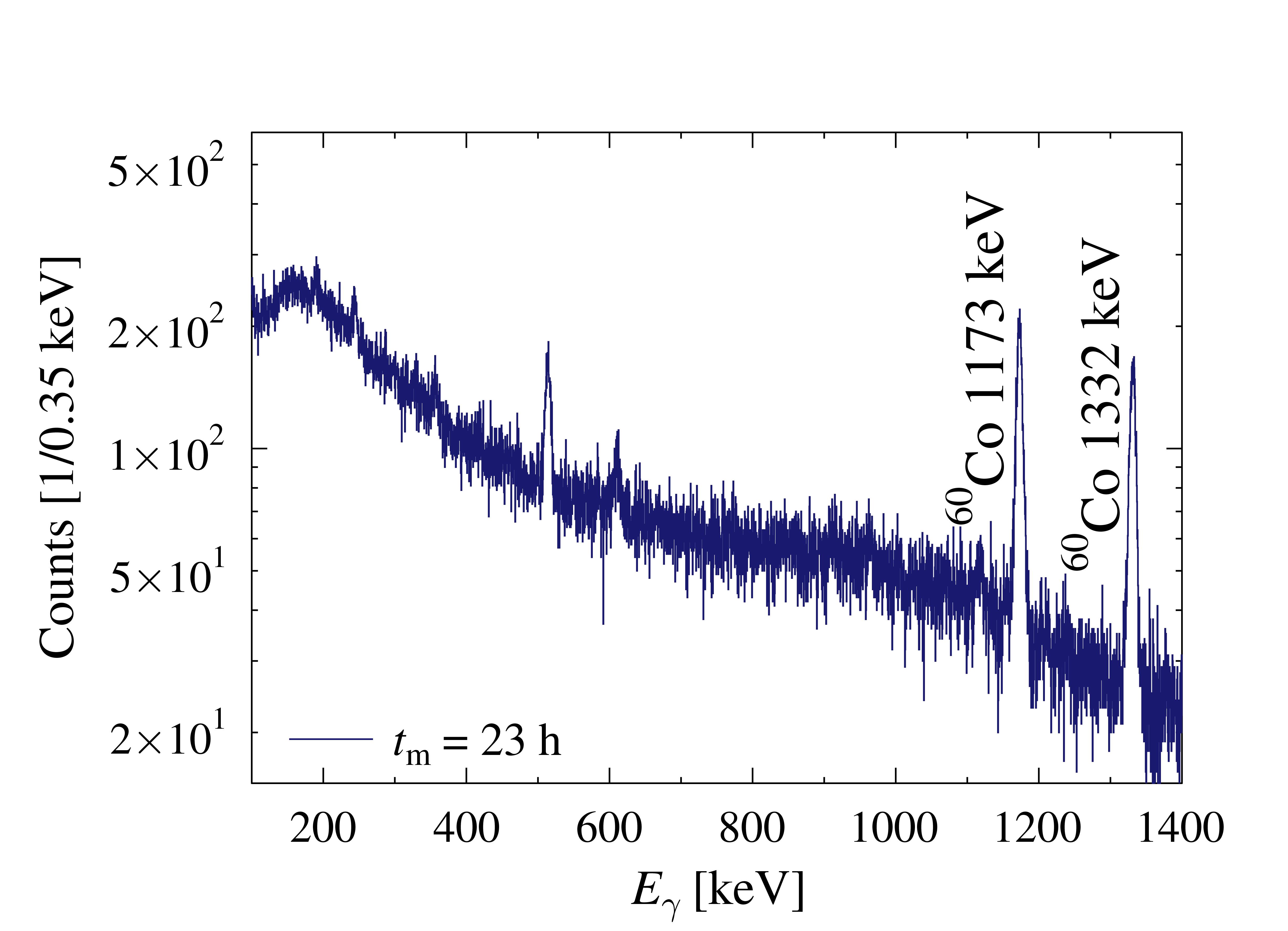}
\caption{(Color online) $\gamma$-ray spectrum for the determination of the $^{60}$Co activity taken 
34 months after purification of the sample. The measurement time was \mbox{$t_{\text{m}}~=~23$ h}.}
\label{sampleParticles}
\end{figure}
The number of $^{60}$Fe atoms in the sample was determined via the increasing 
$^{60}$Co activity \mbox{($t_{1/2}~=~5.272$\,y \cite{Bro13})} according to Fig.~\ref{decay60fe}. The 
activity of $^{60}$Co nuclei increases as 
\begin{equation}
A(^{60}\rm{Co}) = \left( 1-\rm{exp} [-\lambda(^{60}\rm{Co}) t ]\right) A(^{60}\rm{Fe}),
\end{equation}
where $\lambda$ is the decay constant. The related $\gamma$-activity at time $t$ can be derived using the integrated \mbox{line strength $C_\gamma$}

\begin{align}
A_t(^{60}\text{Co}) = &\frac{C_{\gamma}}{0.9976\ \epsilon_{\gamma} \ I_{\gamma}} \nonumber \\
& \ \frac{\lambda}{\mbox{\rm{exp}}\{-\lambda t\}-\mbox{\rm{exp}}\{-\lambda(t+t_{\text{m}})\}},
\end{align}

where $\lambda$ is the decay constant of $^{60}$Co and with the factor $0.9976\pm 0.0003$ \cite{Bro13} for the fraction of $^{60}$Co$^m$, that decays to the ground state of $^{60}$Co, the measurement time $t_\text{m}$, and the detection efficiencies for the \mbox{1173\,keV} and \mbox{1332\,keV} $\gamma$-transitions, respectively. For the analysis of the emission line at the energy of \mbox{1332 keV}, the decay of $^{60}$Co$^\text{m}$ has to be corrected. The decay intensities $I_\gamma$ and efficiencies $\epsilon_{\gamma}$ are listed in \mbox{Table \ref{CalStandard},} which summarizes all decay characteristics adopted in the data analysis. 
With $A=\lambda N$, the number of $^{60}$Fe atoms becomes
\begin{equation}
N(^{60}\rm{Fe}) = \frac{A_t(^{60}\rm{Co})}{1-\rm{exp} \{-\lambda(^{60}\rm{Co}) t \} } \frac{1}{\lambda(^{60}\rm{Fe})}.
\end{equation}

The activity measurement of $^{60}$Co was carried out at the GoetheÐUniversit\"{a}t Frankfurt \mbox{38\,months} after the purification using an HPGe detector of 98\,\% relative efficiency (see Fig.~\ref{sampleParticles}). Background due to the activity of the $^{55}$Fe contamination in the sample was suppressed by a lead foil 1\,mm in thickness. The number of $^{60}$Fe atoms in the sample  
\begin{equation}
 N(^{60}\text{Fe})=(7.77\pm0.11_{\text{\tiny{stat}}}\pm0.42_{\text{\tiny{syst}}} )\times 10^{14} 
\end{equation}
has been determined as a weighted average comprising both $^{60}$Co lines. The systematic uncertainty is determined by the $\gamma$-ray detection efficiency, the decay intensities, and the half-life (see Table \ref{CalStandard}). As the half-life of $^{60}$Fe a value \mbox{$t_{1/2} = (2.60\pm0.05)$\,Myr \cite{WBB15}} was used.

\begin{table}[t]  
   \caption{Number of atoms $N$ (in units of 10$^{19}$) and cross sections 
(in barn) of the neutron fluence monitors $^{94}$Zr and $^{96}$Zr .
   \label{tablemonitor}}
   \begin{ruledtabular}
   \begin{tabular}{ccc}
  						&  $^{94}$Zr	 		&   $^{96}$Zr	 	\\
            \hline 
$N$ 					& $2.7331\pm0.044$	& $0.440\pm 0.014$	\\
$N_{\rm{Cd}}$			& $2.7101\pm0.044$	& $0.437\pm 0.014$	\\
$\sigma_{\rm{th}}$   	& $0.0494\pm0.0017$ 	& $0.0229\pm0.0010$\\
$\sigma_{\rm{RI}}$   	& $0.280\pm0.010$ 		& $5.28\pm0.11$  	\\
\end{tabular}
\end{ruledtabular}
Cross sections were obtained from Ref. \cite{Mug06}.
\end{table}

\subsection{Reactor activations} 

In view of the short half-life of the produced $^{61}$Fe nuclei ($t_{1/2}~=~5.98$\,min \cite{Bha99}), the 
activations at the TRIGA research reactor were performed using a pneumatic transport system between the 
irradiation position and the counting room \cite{EbK00,HET06}.  

The $^{60}$Fe sample was activated for \mbox{$t_\text{a} = 10$ min} with and without cadmium 
foils surrounding the sample in both cases. This so-called cadmium-difference-method allows the distinction
between the thermal neutron capture cross section and the resonance integral, which takes into 
account the epithermal component of the reactor neutron spectrum. The reactor spectrum can 
be described as the sum of a thermal component, \emph{i.e.}, a Maxwell-Boltzmann distribution 
corresponding to $kT=~25.3$\,meV, and an epithermal component following an $1/E$-dependence. 
Due to the very large thermal capture cross section of cadmium, a proper cadmium shielding of 
the sample results in a significantly different response to thermal and epithermal neutrons. In 
the ideal case, all thermal neutrons would be absorbed in the cadmium, while the 
epithermal spectrum remains undisturbed.

\begin{table}[t]  
   \caption{The amount of Zr nuclei produced in the activations and the resulting neutron fluences 
without and with cadmium shielding$^a$.}
   \label{fluences}
   \begin{ruledtabular}
   \begin{tabular}{lcc}
                               	&  without Cd                      	& with Cd \\
                                 \hline
$N(^{95}\text{Zr})$ $^b$	 	& $1.517\pm0.005\pm0.018$	& $0.510\pm0.006\pm0.008$	\\
$N(^{97}\text{Zr})$ $^b$ 		& $1.171\pm0.001\pm0.018$ 	& $1.067\pm0.001\pm0.016$ 	\\
$\Phi_{\rm{th}}$ $^c$	& $8.60\pm0.03\pm0.38$   	& $1.21\pm0.01\pm0.16$		\\
$\Phi_{\rm{epi}}$ $^c$	& $0.467\pm0.002\pm0.014$ 	& $0.458\pm0.005\pm0.014$	\\
\end{tabular}
\end{ruledtabular}
$^a$ Uncertainties are statistical and systematic, respectively.\\
$^b$ In units of $10^{9}$.\\
$^c$ In units of $10^{14}$cm$^{-2}$.
\end{table}

\begin{figure}[b]
\includegraphics[width=0.5\textwidth]{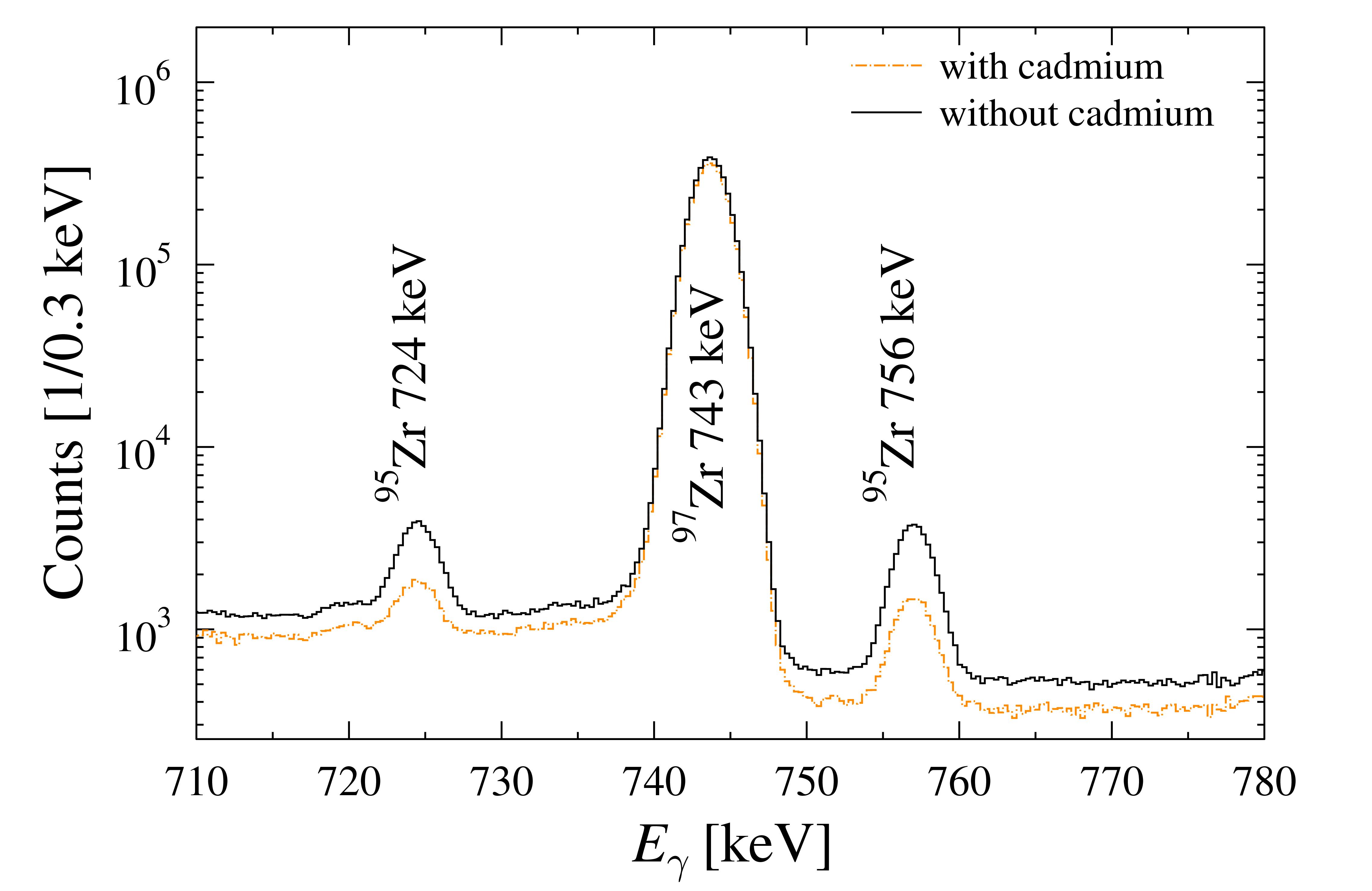}
\caption{(Color online) The $\gamma$-ray spectra of the Zr monitor foils used for the neutron 
fluence analysis normalized for different measuring times. Due to the smaller capture cross section of $^{96}$Zr in the thermal energy regime, the cadmium shielding affects the $^{97}$Zr signature to a much lower degree than that of $^{95}$Zr.}  
\label{Zrfluence}
\end{figure}

The number of product nuclei after the activation $N(^{A+1}X)$ can be expressed in terms of the thermal cross section 
$\sigma_{\text{th}}$, the resonance integral $\sigma_{\text{RI}}$ (\mbox{$\sigma_{\text{RI}} = \int_{E_{\text{\text{cutoff}}}}^{2 \text{MeV}} \sigma(E)/E \text{d}E$} with the cutoff energy \mbox{$E_{cutoff} \approx 90$ meV}), and the epithermal 
($\Phi_{\text{epi}}$) and thermal neutron fluences ($\Phi_{\text{th}}$) in units of cm$^{-2}$,  
\begin{equation}
N(^{A+1}X)  =\ N(^AX) \  (\Phi_{\text{th}} \sigma_{\text{th}} + \Phi_{\text{epi}} \sigma_{\text{RI}}),
\end{equation}
where $N(^{A}X)$ is the number of target nuclei in the irradiated sample. 
\begin{figure*}[t]
\includegraphics[width=1.0\textwidth]{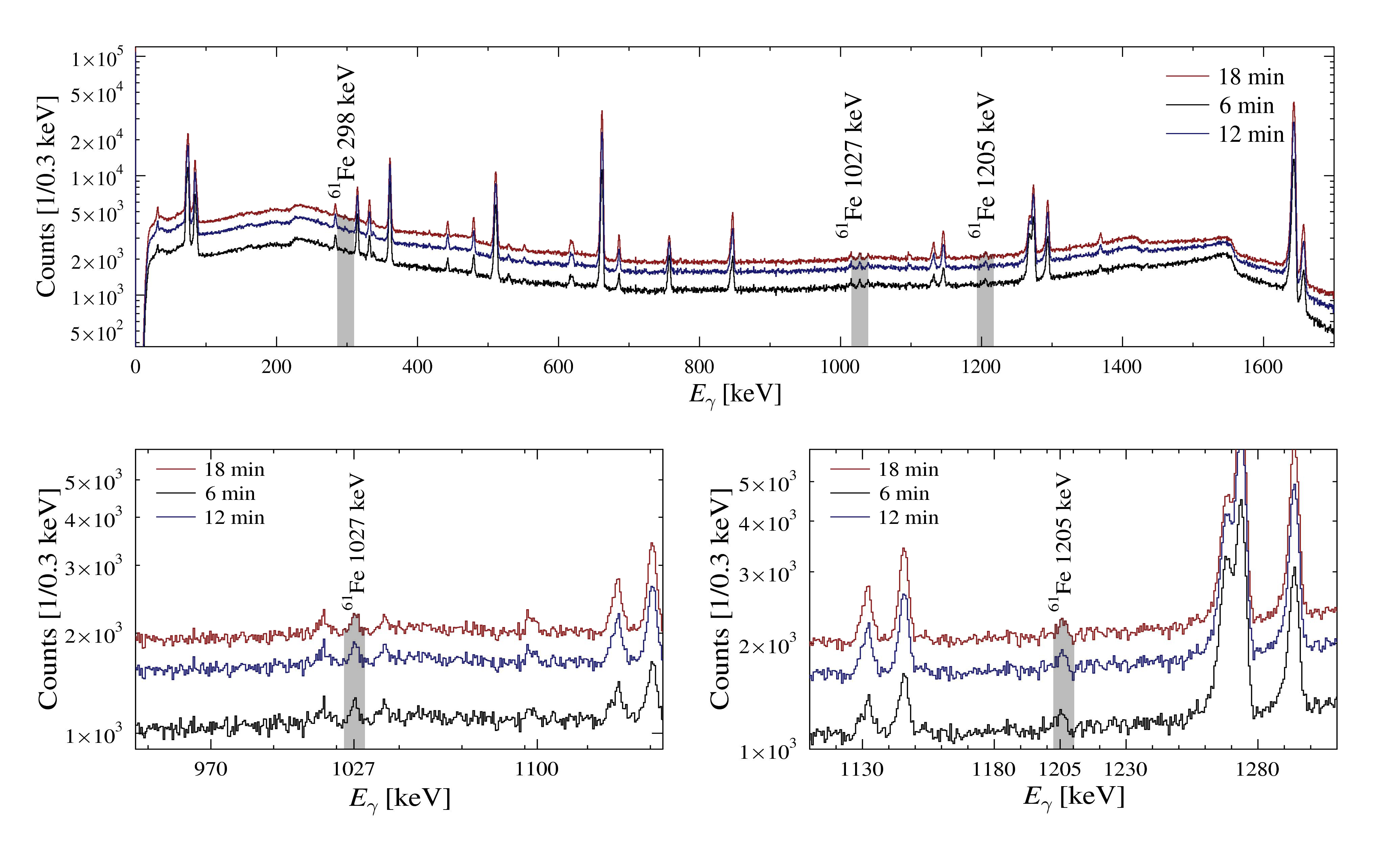}
\caption{(Color online) The $\gamma$-ray spectrum of the activation without cadmium shielding measured for one, two, and three half-lives of $^{61}$Fe, respectively. The upper panel provides an overview, and the lower panels show a zoom into the regions of the $^{61}$Fe lines at 1027\,keV and 1205\,keV. The background stems from the activation of the
contaminants of the $^{60}$Fe sample.}  
\label{61feWithoutCd}
\end{figure*}
Natural zirconium provides a well suited monitor for the epithermal and the thermal 
flux. The activation of the isotopes $^{94}$Zr and $^{96}$Zr exhibit significantly 
different ratios $\sigma_{\text{RI}}/\sigma_{\text{th}}$ (Table \ref{tablemonitor}). The uncertainties in the number of Zr atoms are due to the sample weight (0.2\,\%) and to the isotopic abundances (1.6 and 3.2\,\% for $^{94}$Zr and $^{96}$Zr, respectively) \cite{BeW11}.
Two sets of Zr foils were used in the activations with and without cadmium shielding.



\section{Results}
\subsection{Determination of the neutron fluence}
The zirconium foils used as neutron monitors are 0.127\,mm in thickness and 6\,mm in 
diameter. The foils are thin enough that neutron self-absorption losses during $\gamma$-spectroscopy could be neglected. The fluences in units of 1/cm$^2$ are
\begin{equation}
\Phi_{\text{epi}}=\frac{ N^{97} -  N^{96} \sigma_{\rm{th}}^{96} \Phi_{\rm{th}}}{N^{96} \sigma_{\rm{RI}}^{96}}
\label{eqn:NflussEPI}
\end{equation}
and

\begin{eqnarray*}
\Phi_{\rm{th}} = \frac{N^{96} N^{95} \sigma_{\rm{RI}}^{96} - N^{94} N^{97} \sigma_{\rm{RI}}^{94}}
{N^{94} N^{96}
[\sigma_{\rm{RI}}^{96}\sigma_{\rm{th}}^{94} - \sigma_{\rm{RI}}^{94} \sigma_{\rm{th}}^{96}]},
\label{eqn:NflussTH}
\end{eqnarray*}

where the indices are referring to the various Zr isotopes.
Figure \ref{Zrfluence} shows the $\gamma$-ray spectra of the monitor foils normalized to equal neutron fluence.
Because of the small neutron capture cross section of $^{96}$Zr in the thermal energy regime, the $^{97}$Zr
signal is only marginally affected by the cadmium shielding, whereas $^{95}$Zr exhibits a clear effect 
due to the larger thermal cross section of $^{94}$Zr. The number of produced Zr nuclei is

\begin{equation}
N(^{i}X) = \frac{C_{\gamma}}{\epsilon_{\gamma} I_{\gamma} f_\text{a}  f_\text{w}  f_\text{m}}
\label{ProducedAtoms}
\end{equation}
where
\begin{align}
f_\text{a} = & \frac{1-\exp{(-\lambda_i} t_{\rm a})}{\lambda_i t_{\rm a}},  \\
f_\text{w} = & \exp{(-\lambda_i t_{\rm w})},\\
f_\text{m} = & 1-\exp{(-\lambda_i t_{\rm m})}
\end{align}
are the corrections for the decay during the activation $f_\text{a}$, during the waiting time between activation 
and measurement $f_\text{w}$, and during the measurement $f_\text{m}$, respectively. The correction for the deadtime was of the order of 0.5\,\%. The systematic uncertainty is again determined the \mbox{$\gamma$-efficiencies}, the decay intensities, and the half-life (Table \ref{CalStandard}). The resulting neutron fluences for the two activations are listed in Table~\ref{fluences}. 
\begin{table}[t] 
   \caption{The number of $^{61}$Fe nuclei (in units of 10$^5$) produced in the activations.}
   \label{61FeProd}
   \begin{ruledtabular}
   \begin{tabular}{l l l}
$\gamma$-ray energy             &  \multicolumn{2}{c}{$N$($^{61}$Fe)$^{a}$} \\
	/$\text{keV}$		        & without Cd	& with Cd      	\\
                                 \hline
1027 					&  $1.54\pm0.19\pm0.18$		& $ < 0.179$   	\\
1205 					&  $1.48\pm0.20\pm0.16$		& $ < 0.206$  	\\
 Weighted average 		        & $1.51\pm0.14\pm0.24$  	& $ < 0.179^b $  	\\
\end{tabular}
\end{ruledtabular}
$^a$ Uncertainties are statistical and systematic, respectively.\\
$^b$ Adopted upper limit for further discussion.
\end{table}

\subsection{Thermal ($\text{n}, \gamma$) cross section of $^{60}$Fe}
\begin{figure}[b]
\includegraphics[width=0.5\textwidth]{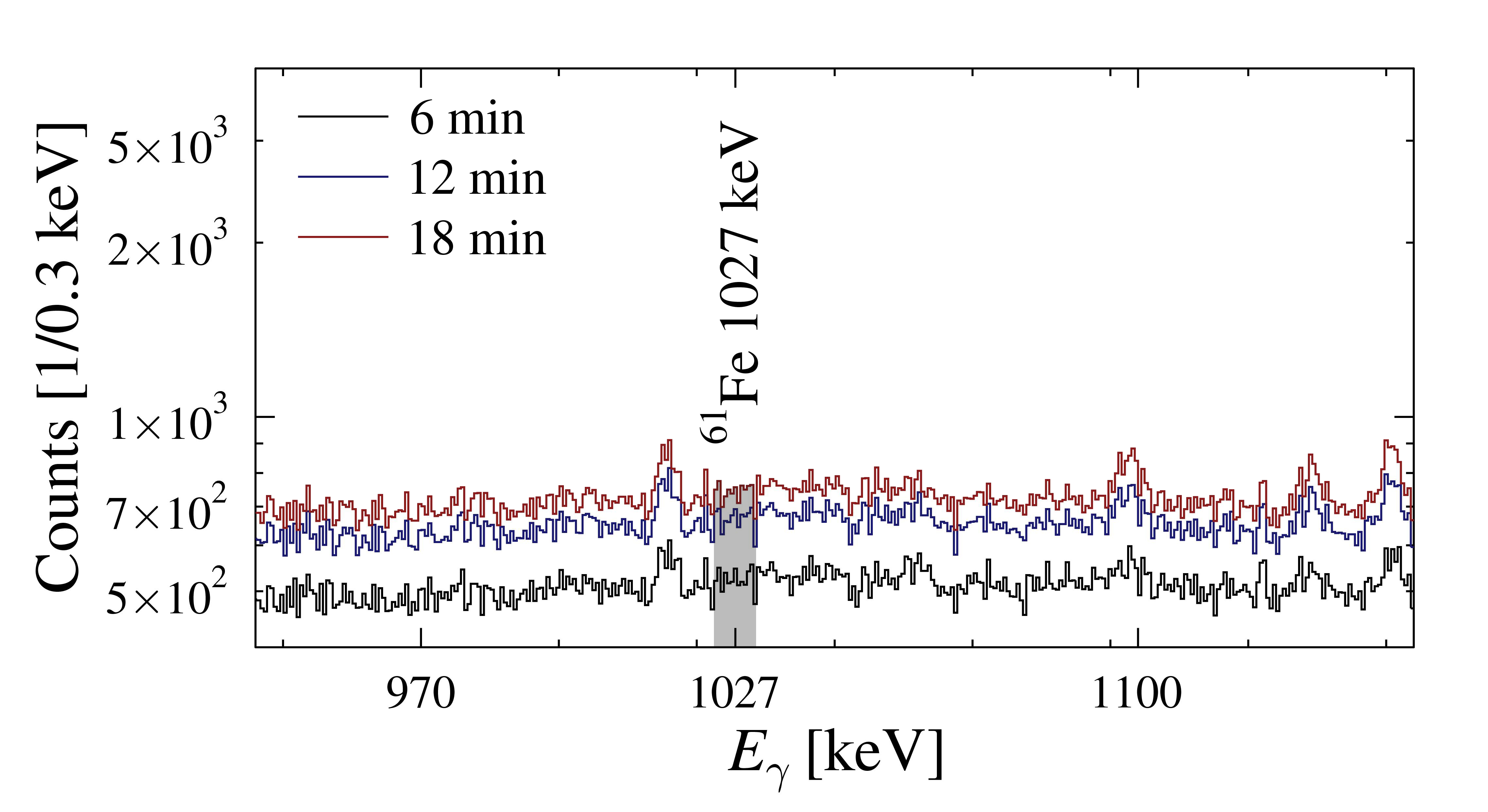}
\caption{(Color online) A detailed view into the region around 1027\,keV of the $\gamma$-ray spectrum after activation with the cadmium 
shielding, illustrating the absence of $^{61}$Fe lines. Therefore, only an upper limit can be deduced
from the data.}  
\label{61feWithCd}
\end{figure}

The $\gamma$-spectrum measured after the activation of the $^{60}$Fe sample without cadmium shielding (Fig.~\ref{61feWithoutCd})
clearly exhibits the $\gamma$-transitions of $^{61}$Fe at 297.9\,keV, 1027\,keV, and 1205\,keV. However, only the last 
two ones were used in the analysis because of the poor signal-to-background ratio of the 298\,keV line. The systematic uncertainty is calculated by the error of the efficiency, the $I_{\gamma}$, the half-lifes, and the neutron fluences (see Table \ref{CalStandard}, Table~\ref{fluences}, and Table \ref{61FeProd}).   \\
In the corresponding spectrum measured after the activation with cadmium shielding, the $^{61}$Fe
lines are completely missing as illustrated in Fig.~\ref{61feWithCd} for the 1027 keV line as an example. In this case, only an upper limit 
can be determined for the resonance integral. The numbers of produced $^{61}$Fe nuclei are listed in 
Table~\ref{61FeProd}.

The number ratio of $^{61}$Fe and $^{60}$Fe after the activation without cadmium is
\begin{equation}
N(^{61}\text{Fe})/N(^{60}\text{Fe}) = \Phi_{\text{th}} \sigma_{\text{th}} + \Phi_{\text{epi}} \sigma_{\text{RI}}.
\label{EgnWithoutCd}
\end{equation}

The thermal cross section 
\begin{equation}
\sigma_{\text{th}}(^{60}\text{Fe}) =\frac{N(^{61}\text{Fe})}{N(^{60}\text{Fe})} \frac{1}{\Phi_{\text{th}}} - \sigma_{\text{RI}} \frac{\Phi_{\text{epi}}}{\Phi_{\text{th}}}
\label{XSth}
\end{equation}
is determined by the number of sample atoms $N(^{60}\text{Fe})$ (Sec.~\ref{SamplePreparation}), the 
neutron fluences $\Phi_{\rm{th}}$ and $\Phi_{\rm{epi}}$ from the Zr monitor measurements (Table~\ref{fluences}), and the number 
of $^{61}$Fe nuclei produced during the activations $N(^{61}\text{Fe})$ (Table~\ref{61FeProd}). The resonance 
integral   
\begin{equation}
\sigma_{\text{RI}} = 
\frac{
       \frac{N^{\text{Cd}}(^{61}\text{Fe})} {N(^{60}\text{Fe})}
      -\frac{\Phi_{\text{th}}^{\text{Cd}}} {\Phi_{\text{th}}} 
       \frac{N(^{61}\text{Fe})} {N(^{60}\text{Fe})}
        }
        {
         \Phi_{\text{epi}}^{\text{Cd}}
        -\frac{\Phi_{\text{th}}^{\text{Cd}} \Phi_{\text{epi}}}
        {\Phi_{\text{th}}}
        },
\label{Ires}
\end{equation}

\begin{figure*}[t]
\includegraphics[width=1.0\textwidth]{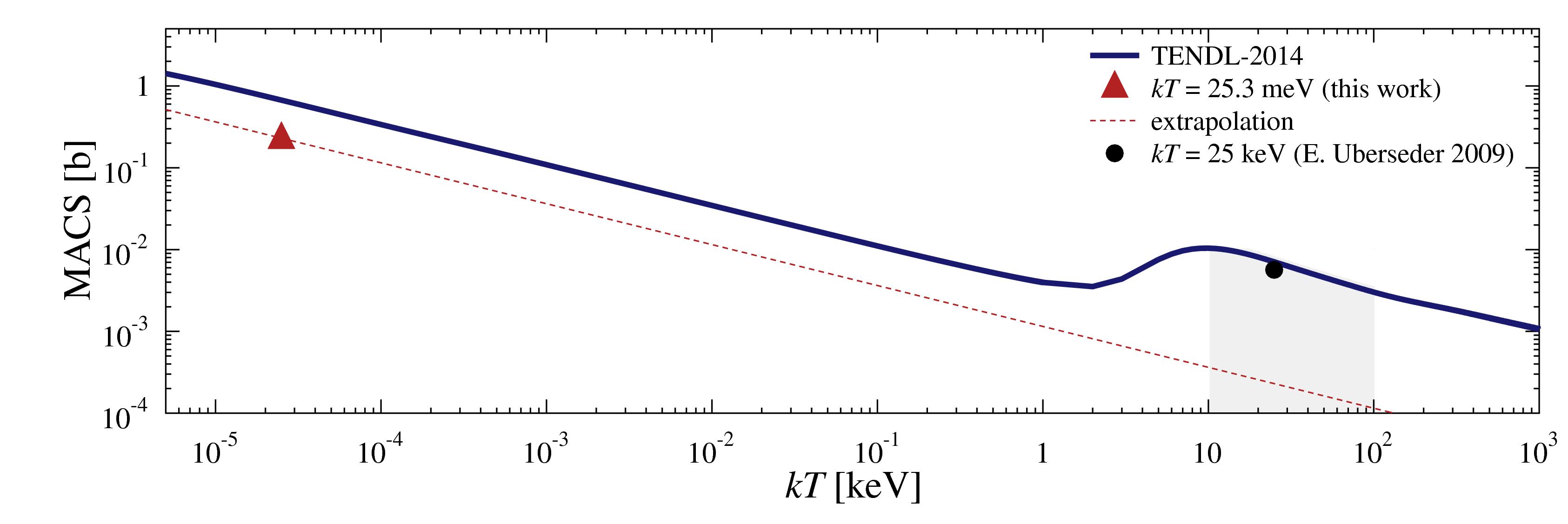}
\caption{(Color online) The Maxwellian averaged cross section for $^{60}$Fe(n,$\gamma$). The present measurement of the cross section at $kT$~=~25.3\,meV (red triangle) and an $0.00115/\sqrt{E}$ extrapolation to the astrophysical energy regime are indicated by the solid red triangle and the dashed red line, respectively. This extrapolation can be used to estimate the DC of the MACS at $kT$~=~25\,keV (black dot) \cite{URS09}. The astrophysical energy regime from $kT$~=~10\,keV~-~100\,keV (grey box) is clearly dominated by the resonant capture contribution. Below about 1\,keV, the MACS based on the most recent version of the TENDL library (TENDL-2014 \cite{KoR12}, blue line) are a factor of 3 above the current measurement. This indicates that the DC componenent is clearly overestimated in the library. }
\label{fig:comparison}
\end{figure*}

is obtained accordingly. Since the epithermal fluences were almost equal in both 
activations, and because the number of $^{61}$Fe nuclei produced with the cadmium absorber 
is much smaller than without absorber, an upper limit for the resonance integral can be defined as
\begin{equation}
\sigma_{\text{RI}} < \frac{N^{\text{Cd}}(^{61}\text{Fe})}{N(^{60}\text{Fe})}\frac{1}{\Phi_{\text{epi}}^{\text{Cd}}}.
\label{epiXS}
\end{equation}
Assuming a 1$\sigma$ confidence level as a constraint
for the resonance integral derived from the 1027\,keV line, one finds
\begin{equation}
0 < \sigma_{\text{RI}} < 0.50\ \text{b}
\label{ri_simple_limit}
\end{equation}
for calculating the thermal cross section using Eq. (\ref{XSth}).

A variation of the resonance integral within these limits affects the thermal cross section  
by about 10\,\%. We assume the resonance integral here explicitly as
\begin{equation}
\sigma_{\text{RI}} = 0.00 ^{+0.50}_{-0.00}\ \text{b},
\label{ri_nice_limit}
\end{equation}
consistent with Eq.~(\ref{ri_simple_limit}), and treat this range as a systematic uncertainty. 
Should the resonance integral be improved in the future, the thermal cross section can be re-evaluated
accordingly. With Eqs.~(\ref{XSth}) and (\ref{ri_nice_limit}) the thermal cross section of $^{60}$Fe 
becomes 
\begin{equation}
\sigma_{\text{th}}(^{60}\text{Fe})
        = (0.226 \pm 0.021_{\text{{stat}}} (^{+0.039}_{-0.045})_{\text{syst}} )   \ \text{b}.
\end{equation}

\section{Summary and Discussion}
Within this work, we characterized the $^{60}$Fe sample to contain $N(^{60}\text{Fe})=(7.77\pm0.11_{\text{\tiny{stat}}}\pm0.42_{\text{\tiny{syst}}} )\times  10^{14}$ atoms. Using the cadmium-difference-method two activations of that sample have been performed 
at the TRIGA research reactor at Johannes Gutenberg-Universit\"{a}t Mainz, Germany. The neutron capture cross section of $^{60}$Fe at 
thermal energies and an experimental upper limit for the resonance integral could be determined 
for the first time:
$$\sigma_{\text{th}}(^{60}\text{Fe}) 
			 = 0.226 \ (^{+0.044}_{-0.049}) \ \text{b} $$

and
$$\sigma_{\text{RI}} < 0.50 \ {\rm b}.$$

Figure~\ref{fig:comparison} shows a comparison of our data with evaluated cross sections 
(TENDL-2014 \cite{KoR12}) and the so far only experimental value of $5.7\pm 1.4$ mb at $kT~=~25\,$keV \cite{URS09,kadonis2009}. Under the assumption that the MACS in the meV-regime is dominated by an s-wave direct capture component, an extrapolation towards higher energies via $1/\sqrt{E}$ is possible. Together with the measurement of the total capture cross section at $kT$=~25\,keV, it is then possible to disentangle the direct and the resonant contribution in the astrophysically interesting energy regime. It turns out that the direct capture component is almost negligible, ranging from less than 10\,\% to less than 1\,\% between 10\,keV and 100\,keV. The comparison of the experimental data with the latest release of TENDL indicates that the resonant component is well described, but the direct capture component is overestimated.

\begin{acknowledgments}
We are very grateful for the excellent support by the entire team of the TRIGA reactor in Mainz. This work was supported by the Helmholtz Young Investigator project VH-NG-327, the BMBF project 05P12RFFN6, the Helmholtz International Center for FAIR and HGS-HIRe. K.S. acknowledges support by DFG \mbox{(SO907/2-1)}. C.L. acknowledges support from the Austrian Science Fund (FWF): J3503.
\end{acknowledgments}

\input{60fe_thermalActivation.bbl}

\end{document}

%% file: 60fe_thermalActivation.bbl
\newcommand{\noopsort}[1]{} \newcommand{\printfirst}[2]{#1}
  \newcommand{\singleletter}[1]{#1} \newcommand{\swithchargs}[2]{#2#1}